\documentclass[a4paper]{jpconf}                                                                                                               
                                                                                                                    
\usepackage{amsmath}                                                                                                                   
\usepackage{amssymb}                                                                                                                        
\usepackage{graphicx}                                                                                                             

\newcommand{\ud}{\mathrm{d}}                                                                                                              
         
\begin{document}

\title{En route to fusion: confinement state as a waveleton}

\author{Antonina N. Fedorova, Michael G. Zeitlin}

\address{IPME RAS, St.~Petersburg,
V.O. Bolshoj pr., 61, 199178, Russia\\
http://math.ipme.ru/zeitlin.html, http://mp.ipme.ru/zeitlin.html}

\ead{zeitlin@math.ipme.ru, anton@math.ipme.ru}

\begin{abstract}
A fast and efficient numerical-analytical approach is proposed for
description of complex behaviour in non-equilibrium ensembles in the
BBGKY framework. We construct the multiscale representation for
hierarchy of partition functions by means of the variational approach
and multiresolution decomposition. Numerical modeling shows the creation
of various internal structures from fundamental localized (eigen)modes.
These patterns determine the behaviour of plasma. The localized pattern
(waveleton) is a model for energy confinement state (fusion) in plasma.
\end{abstract}

\begin{flushright}
{\bf ``A magnetically confined plasma cannot \\
be in thermodinamical equilibrium'' }\\
Unknown author ... Folklore \\
\end{flushright}

\section{Introduction}
It is well known that fusion problem in plasma physics could be solved neither 
experimentally nor theoretically during last fifty year, so it seems that there are the serious
obstacles which prevent real progress in the problem of real fusion as
the main subject in the area [1], [2]. Of course, it may be a result of some
unknown no-go theorem(s) but it seems that the current theoretical level
demonstrates that not all possibilities, at least on the
level of theoretical and matematical modeling, are exhausted. 
Definitely, the first thing which we need to change is a framework of
generic mathematical methods which can help to improve the current state of
the theory. 
Our postulates (conjectures) are as follows [3]: 

{\bf A)} The fusion problem (at least at the first step) must be considered as
a problem inside the (non) equilibrium ensemble in the full phase space.
It means, at least, that: 

{\bf A1)} our dynamical variables are partitions (partition functions,
hierarchy of N-points partition functions), 

{\bf A2)} it is impossible to fix a priori the concrete distribution function
and postulate it (e.g. Maxwell-like or other concrete (gaussian-like or even not) distributions)
but, on the contrary, the proper distrubution(s) must be the solutions
of proper (stochastic) dynamical problem(s), e.g., it may be the
well-known framework of BBGKY hierarchy of kinetic equations or
something similar. So, the full set of dynamical variables must include
partitions also. 

{\bf B)} Fusion state = (meta) stable state (with minimum entropy and zero measure) 
in the space of partitions on the
whole phase space in which most of energy of the system is concentrated
in the relatively small area (preferable with measure zero) of the whole
domain of definition in the phase space during the time period which is
enough to take reasonable part of it outside for possible usage. 
From the formal/mathematical point of view it means that: 

{\bf B1)} fusion state must be localized (first of all, in the phase space), 

{\bf B2)} we need a set of building blocks, localized basic states or
eigenmodes which can provide 

{\bf B3)} the creation of localized pattern which can be considered as a
possible model for plasma in a fusion state. Such pattern must be: 

{\bf B4)} (meta) stable and controllable, because of obvious reasons. 
So, the main courses are: 

{\bf C1)} to present smart localized building blocks which may be not only
useful from point of view of analytical statements, such as the best
possible localization, fast convergence, sparse operators representation,
etc, but also exist as real physical fundamental modes, 

{\bf C2)} to construct various possible patterns with special attention to
localized pattern which can be considered as a needful thing in analysis
of fusion; 

{\bf C3)} after points C1 and C2 in ensemble (BBGKY) framework to consider
some standard reductions to Vlasov-like and RMS-like equations
(following the set-up from well-known results) which may be useful
also. These particular cases may be important as from physical point of
view as some illustration of general consideration.

The lines above are motivated by our attempts to analyze the hidden
internal contents of the phrase mentioned in the epigraph of this paper:
``A magnetically confined plasma cannot be in thermodinamical
equilibrium.'' Also, it should be noted that our results below can be applied to 
any scenario (fusion, ignition, etc): we describe pattern formation in arbitrary 
non-equilibrium ensembles. 


\section{Description}

At this stage our main goal is an attempt of classification and
construction of a possible zoo of nontrivial (meta) stable
states/patterns: high-localized (nonlinear) eigenmodes, complex
(chaotic-like or entangled) patterns, localized (stable) patterns
(waveletons). We will use it later for fusion description, modeling and
control. 
In our opinion localized (meta)stable pattern (waveleton) is the proper image for fusion
state in plasma (energy confinement). 

Our constructions can be applied 
to the hierarchy of  $N$-particle distribution function, satisfying the standard BBGKY 
hierarchy ($\upsilon$ is the volume): 
\begin{equation}
\frac{\partial F_s}{\partial t}+L_sF_s=
\frac{1}{\upsilon}\int d\mu_{s+1}
\sum^s_{i=1}L_{i,s+1}F_{s+1}.
\end{equation}
Our key point is the proper nonperturbative generalization of the
previous perturbative multiscale approaches (like Bogolubov/virial expansions).
The infinite hierarchy of
distribution functions is: 
{\setlength\arraycolsep{0pt}
\begin{eqnarray}
&&F=\{F_0,F_1(x_1;t),\dots,
F_N(x_1,\dots,x_N;t),\dots\},\nonumber\\
&&F_p(x_1,\dots, x_p;t)\in H^p,\quad
H^0=R,\ H^p=L^2(R^{6p}),\nonumber\\
&&F\in H^\infty=H^0\oplus H^1\oplus\dots\oplus H^p\oplus\dots
\end{eqnarray}}
with the natural Fock space like norm (guaranteeing the positivity of
the full measure): 
\begin{equation}
(F,F)=F^2_0+\sum_{i}\int F^2_i(x_1,\dots,x_i;t)\prod^i_{\ell=1}\mu_\ell.
\end{equation}
Multiresolution decomposition (filtration) naturally and efficiently
introduces the infinite sequence (tower) 
of the underlying hidden scales, which is a sequence of increasing closed subspaces
$V_j\in L^2(R)$ [4]:
\begin{equation}
...V_{-2}\subset V_{-1}\subset V_0\subset V_{1}\subset V_{2}\subset ...
\end{equation}
Our variational approach [3] reduces the initial problem to the problem of
solution of functional equations at the first stage and some algebraic
problems at the second one.
Let $L$ be an arbitrary (non)li\-ne\-ar dif\-fe\-ren\-ti\-al\-/\-in\-teg\-ral operator 
 with matrix dimension $d$
(finite or infinite), 
which acts on some set of functions
from $L^2(\Omega^{\otimes^n})$:  
$\quad\Psi\equiv\Psi(t,x_1,x_2,\dots)=\Big(\Psi^1(t,x_1,x_2,\dots), \dots$,
$\Psi^d(t,x_1,x_2,\dots)\Big)$,
 $\quad x_i\in\Omega\subset{\bf R}^6$, $n$
 is the number of particles:  
{\setlength\arraycolsep{0pt}
\begin{eqnarray}
L\Psi&\equiv& L(Q,t,x_i)\Psi(t,x_i)=0,\nonumber\\
Q&\equiv& Q_{d_0,d_1,d_2,\dots}(t,x_1,x_2,\dots,
\partial /\partial t,\partial /\partial x_1,
\partial /\partial x_2,
\dots,\int \mu_k)\nonumber\\
&=&
\sum_{i_0,i_1,i_2,\dots=1}^{d_0,d_1,d_2,\dots}
q_{i_0i_1i_2\dots}(t,x_1,x_2,\dots)
\Big(\frac{\partial}{\partial t}
\Big)^{i_0}\Big(\frac{\partial}{\partial x_1}\Big)^{i_1}
\Big(\frac{\partial}{\partial x_2}\Big)^{i_2}\dots\int\mu_k.
\end{eqnarray}			
}
\noindent Let us consider now the $N$		
mode approximation for the solution as the following ansatz: 
{\setlength\arraycolsep{0pt}
\begin{eqnarray}
\Psi^N(t,x_1,x_2,\dots)=
\sum^N_{i_0,i_1,i_2,\dots=1}a_{i_0i_1i_2\dots}
 A_{i_0}\otimes 
B_{i_1}\otimes C_{i_2}\dots(t,x_1,x_2,\dots).
\end{eqnarray}
}
We shall determine the expansion coefficients from the following
conditions: 
{\setlength\arraycolsep{0pt}
\begin{eqnarray}
\ell^N_{k_0,k_1,k_2,\dots}\equiv 
\int(L\Psi^N)A_{k_0}(t)B_{k_1}(x_1)
C_{k_2}(x_2) \ud t \ud x_1 \ud x_2 \dots=0.
\end{eqnarray}}
As a result the solution has the following multiscale/multiresolution
decomposition via nonlinear high-localized eigenmodes [4]:
{\setlength\arraycolsep{-2pt}
\begin{eqnarray}
&&F(t,x_1,x_2,\dots)=\sum_{(i,j)\in Z^2}a_{ij}U^i\otimes V^j(t,x_1,x_2,\dots),\nonumber \\
&&V^j(t)=V_N^{j,slow}(t)+\sum_{l\geq N}V^j_l(\omega_lt), \ \omega_l\sim 2^l, \\
&&U^i(x_s)=U_M^{i,slow}(x_s)+\sum_{m\geq M}U^i_m(k^{s}_mx_s), \ k^{s}_m\sim 2^m.
 \nonumber
\end{eqnarray}
}
\noindent 
So, we may move from the coarse scales
of resolution (coarse graining) to the finest ones for obtaining more detailed information
about the dynamical process. 
In this way one obtains contributions to the full solution from each
scale of resolution or each time/space scale or from each nonlinear
eigenmode. 
It should be noted that such representations give the best possible
localization properties in the corresponding (phase) space/time
coordinates. 
Numerical calculations are based on compactly supported wavelets and related wavelet 
families and on evaluation of the accuracy on the level $N$
 of the corresponding cut-off of the full system w.r.t. the norm (3):\\
\begin{eqnarray}
\|F^{N+1}-F^{N}\|\leq\varepsilon
\end{eqnarray}
Numerical modeling shows the creation of various complex structures
from localized modes, which are related to (meta)stable or unstable type
of behaviour and the corresponding patterns (waveletons) formation (Figs.~1, 2).
Reduced algebraic structure (7), Generalized Dispersion Relations, 
provide the pure algebraic control of
stability/unstability scenario. 
So, we considered the construction for 
controllable (meta) stable waveleton configuration representing a
reasonable approximation for the possible realizable confinement state.

\section{Conclusions}
Let us summarize our main results: 

{\bf Physical Conjectures:} 

\noindent{\bf P1} 
State of fusion (confinement of energy) in plasma physics may and need
be considered from the point of view of non-equilibrium statistical
physics. According to this BBGKY framework looks naturally as first
iteration. Main dynamical variables are partitions.

\noindent{\bf P2} 
Basic high localized nonlinear eigenmodes are real physical
modes important for fusion modeling. Intermode multiscale interactions
create various complex patterns from these fundamental building blocks, and
determine the behaviour of plasma (Figs.~1, 2). 
High localized (meta) stable patterns (waveletons), considered as
long-living fluctuations, are proper images for plasma in fusion state
(Fig.~2). 

{\bf Mathematical framework:} 

\noindent{\bf M1} 
The problems under consideration, like BBGKY hierarchies or their reductions 
are considered as pseudodifferential hierarchies in the 
framework of proper family of methods unified by effective multiresolution approach
or local nonlinear harmonic analysis on the orbits of representations of hidden
underlying symmetry of properly chosen functional space [3].

\begin{figure}[h]
\begin{minipage}{18pc}
\includegraphics[width=18pc]{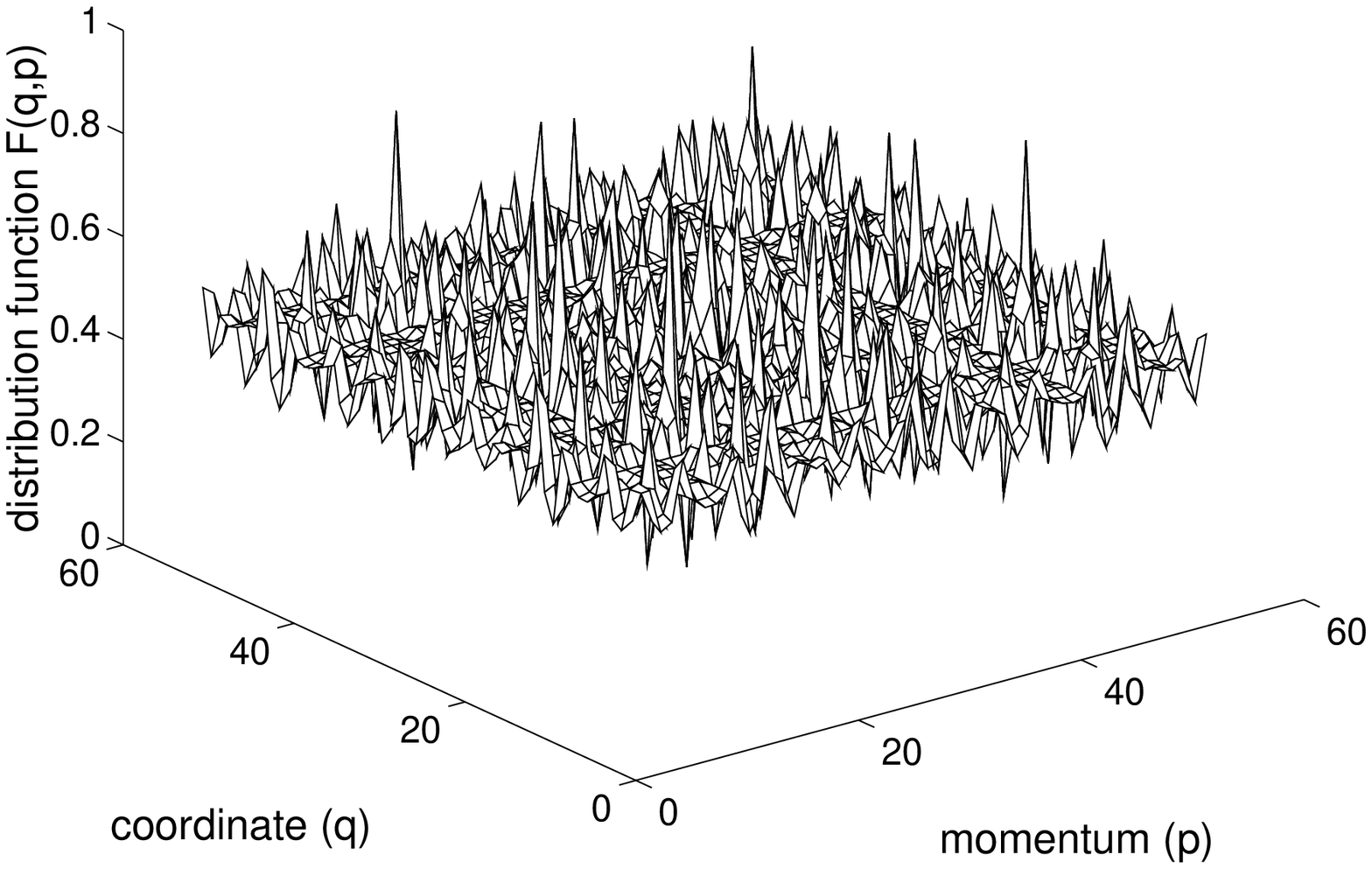}
\caption{Trash: Chaotic Partition.}
\end{minipage}\hspace{2pc}%
\begin{minipage}{18pc}
\includegraphics[width=18pc]{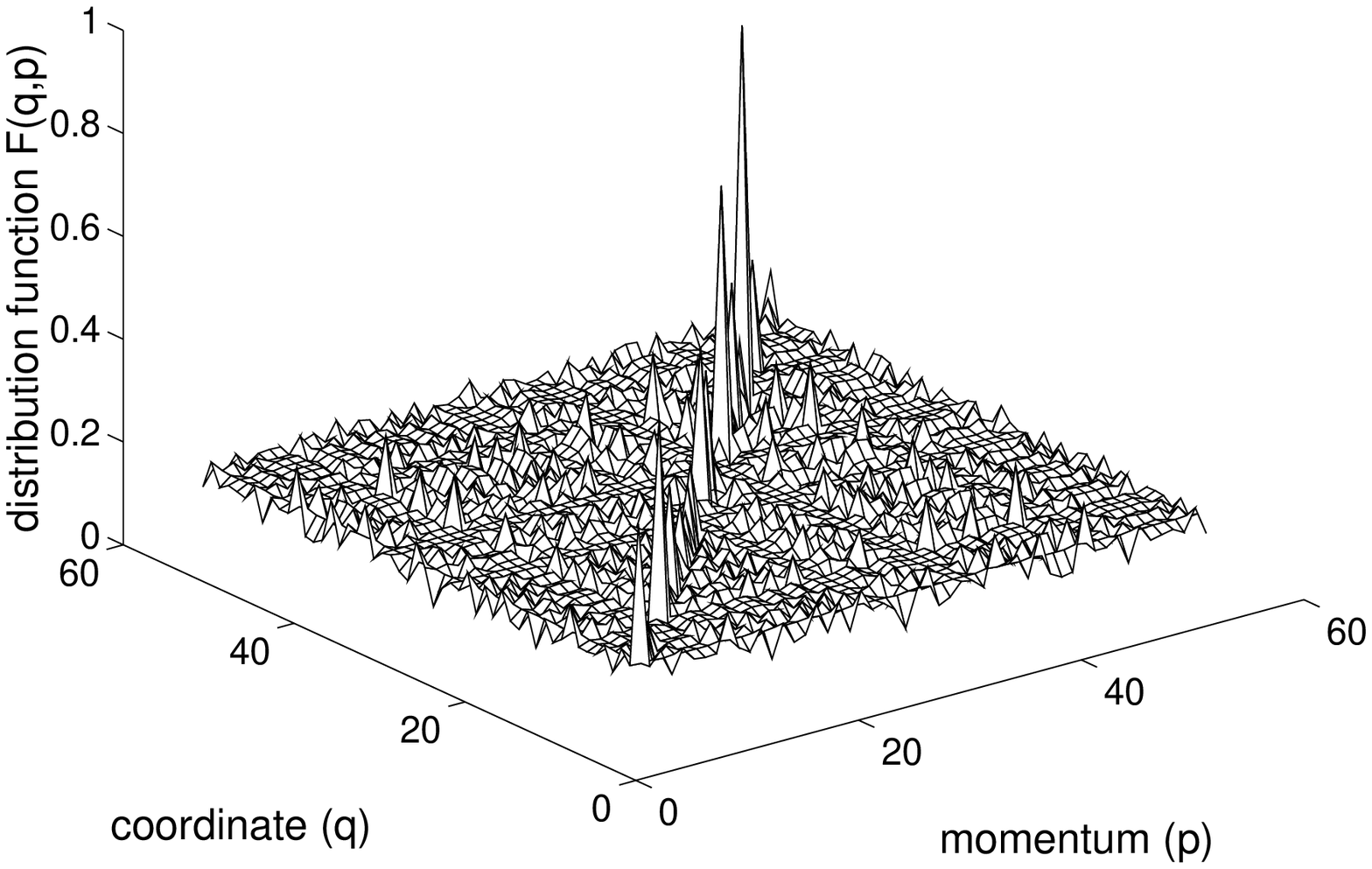}
\caption{Goal: Waveleton/Fusion State. }
\end{minipage}
\end{figure}

\noindent{\bf M2} 
Formulas (8) based on Generalized Dispersion Relations (GDR) (7)
provide exact multiscale representation for all dynamical variables
(partitions, first of all) in the basis of high-localized nonlinear
(eigen)modes. Numerical realizations in this framework are maximally
effective from the point of view of complexity of all algorithms inside.
GDR provide the way for the state control on the pure algebraical level. 

\noindent{\bf Realizability:} 

According to this approach, it is possible on formal level, in
principle, to control ensemble behaviour and to realize the localization
of energy (confinement state) inside the waveleton configurations
created from a few fundamental modes only during 
self-organization via possible (external) algebraical control (Figs.~1, 2) [3].


\end{document}